\documentclass[sn-mathphys-num]{sn-jnl}

\usepackage{graphicx}%
\usepackage{multirow}%
\usepackage{amsmath,amssymb,amsfonts}%
\usepackage{amsthm}%
\usepackage{mathrsfs}%
\usepackage[title]{appendix}%
\usepackage{xcolor}%
\usepackage{textcomp}%
\usepackage{manyfoot}%
\usepackage{booktabs}%
\usepackage{algorithm}%
\usepackage{algorithmicx}%
\usepackage{algpseudocode}%
\usepackage{listings}%
\usepackage{bm}%

\theoremstyle{thmstyleone}%
\theoremstyle{thmstyletwo}%

\theoremstyle{thmstylethree}%

\begin{document}

\title[Triple-core structure of the double-core vortex in $^3$He-B]{Triple-core structure of the double-core vortex in superfluid $^3$He-B}


\author{\fnm{Riku} \sur{Rantanen}}\email{riku.s.rantanen@aalto.fi}

\affil{\orgdiv{Department of Applied Physics}, \orgname{Aalto University}, \orgaddress{PO Box 15100, FI-00076 AALTO, Finland}}


\abstract{The order parameter of superfluid $^3$He involves nine complex components, and the multicomponent structure allows quantized vortices in superfluid $^3$He to have complicated cores. One of the vortices found in the B phase is the double-core vortex, which has been often described as a pair of two half-quantum vortices (HQVs) connected by a domain wall. Our numerical calculations of the core structure suggest an alternative representation of the vortex as a combination of three vortices, one in each component of the spin-triplet superfluid. Based on the results we present a qualitative analytical model for the triple-core structure of the double-core vortex. Additionally we numerically calculate the structure of a double-core vortex stretched between pinning sites, and show that the HQV picture becomes more applicable when separation between subcores becomes large.}




\maketitle

\section{Introduction}\label{sec1}

Superfluid helium-3 is currently the only superfluid (or superconducting) system with spin-triplet pairing where the form of the order parameter is explicitly known. Compared to a simple s-wave superconductor, the multicomponent order parameter leads to increased complexity in practically all features of the superfluid, but in particular the structure of quantized vortices can be very different. The major qualitative difference is that superfluidity is not necessarily completely suppressed in the cores of vortices, but instead the cores are filled with a superfluid component distinct from the bulk. In the bulk, the state of the order parameter is defined purely by the bulk energy, while near the vortex core the kinetic energy from phase winding is the most important contribution. Another feature of the spin-triplet superfluid is that in addition to the typical mass superflow, spin superfluidity is also possible which leads to the existence of spin vortices. 

In the B phase of superfluid $^3$He, two single-quantum mass vortices are known to exist~\cite{Thuneberg1987,Salomaa1987,Regan2020,Rantanen2024}: the axially symmetric A-phase-core vortex and the asymmetric double-core vortex. In this article, we focus on the structure of the double-core vortex, which is typically thought of as a tightly bound pair of half-quantum vortices (HQVs). Based on numerical results, we present a simple analytic form for the order parameter near the vortex axis that represents the double-core vortex as a group of three vortices, one in each of the three spin components of the B phase. A vortex split into three subcores was recently shown to also be possible in thin films of $^3$He-A~\cite{Gluscevich2025}.

The order parameter of superfluid $^3$He is described in Section~\ref{sec:orderparameter}, along with its representation in the angular momentum basis. In Section~\ref{sec:polarhqv} we discuss the structure of HQVs in the polar phase as an introduction to the structure of vortices in $^3$He. The analytic form of the double-core vortex is presented in Section~\ref{sec:doublecore} and numerical calculation of the core structure shows that the division to three separate cores is accurate in the bulk case. We also calculate the vortex state numerically in a situation where the the two subcores of the vortex are pinned far apart. When the separation between subcores becomes large, the description of two HQVs connected by a domain wall becomes more accurate. A very similar structure is predicted to be the state found in the polar-distorted B phase in anisotropically confined superfluid $^3$He~\cite{Makinen2019}. The final section includes our conclusions.

\section{The order parameter of superfluid $^3$He}\label{sec:orderparameter}

Superfluid $^3$He consists of Cooper pairs in a spin-triplet p-wave state. The order parameter of superfluid $^3$He describes the internal state of these Cooper pairs. The $2\times 2$ spin-triplet p-wave pairing gap matrix is
\begin{equation}
    \Delta(\hat{\bm{k}}) = \sum_{\mu j} A_{\mu j}\hat{k}_j i\bm{\sigma}_\mu \sigma_y = \begin{bmatrix}
        \Delta_{\uparrow\uparrow} & \Delta_{\uparrow\downarrow} \\
        \Delta_{\downarrow\uparrow} & \Delta_{\downarrow\downarrow}
    \end{bmatrix}
    \label{eq:gapmatrix}
\end{equation}
where $\bm{\sigma} = (\sigma_x, \sigma_y, \sigma_z)$ are the Pauli matrices, $\hat{\bm{k}}$ is the direction in momentum space and the subscript arrows indicate the spin state of the Cooper pairs. The gap matrix is parametrized by a $3\times 3$ complex matrix $A_{\mu j}$, commonly referred to as the order parameter, with the first and second indices corresponding to degrees of freedom in spin and orbital spaces, respectively.

Typically the order parameter is written in a Cartesian basis, so that $\mu, j = x,y,z$ correspond to directions in their respective spaces. Alternatively the order parameter can be expressed in the basis of angular momentum eigenvectors~\cite{Ohmi1983} $\hat{\bm{e}}^\pm=(\hat{\bm{x}}\pm i\hat{\bm{y}})/\sqrt{2}$ and $\hat{\bm{e}}^0=\hat{\bm{z}}$, corresponding to the possible spin and orbital angular momentum projections of the Cooper pairs: $\pm1$ and $0$. Note that here the quantization axis has been chosen along $\hat{\bm{z}}$. Converting between the two representations can be done with a change of basis
\begin{equation}
    A' = U A U^T,\quad U = \left[\hat{\bm{e}}^-\:\hat{\bm{e}}^0\:\hat{\bm{e}}^+\right]^T
\end{equation}
where $A'_{\kappa \nu}$ is the order parameter in the angular momentum basis, with $\kappa, \nu = -1, 0, +1$ corresponding to the spin and angular momentum projections $|S_z, L_z\rangle = |\kappa, \nu\rangle$ of the Cooper pairs. For clarity, we refer to the angular momentum basis components by the amplitude $|A'_{\kappa\nu}| = C_{\kappa\nu}$ and phase $\phi_{\kappa\nu}$ of the individual order parameter components separately, so that $A'_{\kappa \nu} = C_{\kappa\nu}e^{i\phi_{\kappa\nu}}$

Depending on the situation, the $C_{\kappa \nu}$ representation can often be more physically intuitive than its Cartesian counterpart, as it describes the state of the Cooper pairs directly. For example, the B phase order parameter in Cartesian basis can be expressed as
\begin{equation}
    A_{\mu j} = \Delta e^{i\phi}R(\bm{\theta}) = \Delta e^{i\phi}(\hat{d}_\mu \hat{m}_j + \hat{e}_\mu \hat{n}_j + \hat{f}_\mu \hat{l}_j)
    \label{eq:bphase1}
\end{equation}
where $\Delta$ is the amplitude and $\phi$ is the phase the order parameter, and the rotation matrix $R(\bm{\theta})$ describes the relative orientation of the spin and orbital spaces as a rotation by $\theta = |\bm{\theta}|$ around an axis $\hat{\bm{\theta}} = \bm{\theta}/\theta$. The vectors $(\bm{\hat{d}},\hat{\bm{e}},\bm{\hat{f}})$ and $(\bm{\hat{m}},\bm{\hat{n}},\bm{\hat{l}})$ form the coordinate basis for the spin and orbital spaces, respectively. If we choose the quantization axis of our angular momentum representation such that it is along the axis of rotation $\hat{\bm{\theta}}$, we find that the order parameter only has the nonzero components $C_{+-}$, $C_{00}$ and $C_{-+}$, i.e. the B phase only contains Cooper pairs that have total angular momentum $J = S + L = 0$. This division into angular momentum components suggests an alternative way of writing Eq.~\eqref{eq:bphase1}, which we will see is extremely useful when describing the double-core vortex of B phase.

\section{Half-quantum vortices in the polar phase}\label{sec:polarhqv}

Before delving into the B phase, it is helpful to first look at a simpler example of HQVs in $^3$He. The polar phase is an equal-spin pairing state with an order parameter
\begin{equation}
    A_{\mu j} = \Delta e^{i\phi}\hat{d}_\mu \hat{m}_j
    \label{eq:polarphase}
\end{equation}
where $\hat{\bm{d}}$ is a unit vector in the plane perpendicular to the spins of the Cooper pairs and $\hat{\bm{m}}$ is a unit vector in orbital space. Experimentally the polar phase is stabilized by anisotropic confinement, for example with an array of thin strands all aligned along a common axis~\cite{Dmitriev2015}. In this case, $\bm{\hat{m}}$ is strongly locked along the direction of the strands. In the absence of an external magnetic field, the $\hat{\bm{d}}$ vector remains in the plane perpendicular to $\hat{\bm{m}}$ due to spin-orbit coupling. If we choose axes so that $\hat{\bm{m}} = \hat{\bm{z}}$, then the order parameter in Eq.\eqref{eq:polarphase} can be written as
\begin{equation}
    A_{\mu j} = \Delta e^{i\phi} \left(\hat{x}_\mu\cos\theta + \hat{y}_\mu\sin\theta\right)\hat{z}_j
    \label{eq:polarphase2}
\end{equation}
with the angle of $\hat{\bm{d}}$ in the $xy$-plane given by $\theta$.

For the polar phase in the state described by Eq.~\eqref{eq:polarphase2}, choosing $\hat{\bm{z}}$ as the quantization axis we find that the only non-zero components are $C_{+0}$ and $C_{-0}$, the spin up and spin down components with zero orbital angular momentum projection. With this knowledge we can rewrite Eq.~\eqref{eq:polarphase2} as two components each with their own amplitude and phase:
\begin{equation}
    A_{\mu j} = \Delta_+ e^{i\phi_+}(\hat{x}_\mu + i\hat{y}_\mu)\hat{z}_j + \Delta_- e^{i\phi_-}(\hat{x}_\mu - i\hat{y}_\mu)\hat{z}_j
    \label{eq:polarupdown}
\end{equation}
with the subscripts $+$ and $-$ indicating spin up and spin down components, respectively, and $\Delta_+ = C_{+0}$ and $\Delta_- = C_{-0}$. In the bulk polar phase in the absence of a magnetic field $\Delta_+ = \Delta_- = \Delta/2$ and Eqs.~\eqref{eq:polarphase2}, and \eqref{eq:polarupdown} are equivalent:
\begin{align}
    A_{\mu j} &= \Delta_+ e^{i\phi_+}(\hat{x}_\mu + i\hat{y}_\mu)\hat{z}_j + \Delta_- e^{i\phi_-}(\hat{x}_\mu - i\hat{y}_\mu)\hat{z}_j \nonumber \\ 
    &= \frac{\Delta}{2}\left(\hat{x}_\mu(\cos\phi_+ + \cos\phi_-) + \hat{y}_\mu(\sin\phi_- - \sin\phi_+)\right)\hat{z}_j \nonumber \\
    &+ i\frac{\Delta}{2}\left(\hat{x}_\mu (\sin\phi_+ + \sin\phi_-) + \hat{y}_\mu (\cos\phi_+ - \cos\phi_-)\right)\hat{z}_j \nonumber \\
    &= \Delta e^{i(\phi_+ + \phi_-)/2}\left[\cos\left(\frac{\phi_- - \phi_+}{2}\right)\hat{x}_\mu + \sin\left(\frac{\phi_- - \phi_+}{2}\right)\hat{y}_\mu\right]\hat{z}_j \nonumber \\
    &= \Delta e^{i\phi}(\hat{x}_\mu \cos\theta + \hat{y}_\mu \sin\theta)\hat{z}_j
    \label{eq:polarequivalency}
\end{align}
with $\phi = \frac{\phi_+ + \phi_-}{2}$ and $\theta = \frac{\phi_- - \phi_+}{2}$. An equal winding in both components corresponds to a pure phase winding in the polar phase, while an equal but opposite winding corresponds to a rotation of the spin anisotropy vector $\hat{\bm{d}}$ in the plane.

This split into two components allows us to easily understand different vortex structures in the polar phase. Consider the case where the two spin components both have a positive winding vortex centered at the same point, with the corresponding amplitudes suppressed at the winding axis. This is simply the single-quantum vortex in the polar phase, with $\phi$ winding by $2\pi$, uniform $\hat{\bm{d}}$, and fully suppressed superfluidity in the core. When the two components have vortices with opposite windings, we get a spin-vortex with no $\phi$ winding and $\hat{\bm{d}}$ rotating by $2\pi$ in the $xy$ plane.

What about winding in only one component, for example in $\phi_+$? In this case the total phase winds by $\phi = (2\pi + 0)/2 = \pi$, and simultaneously the $\hat{\bm{d}}$ vector rotates by $\theta = (0 - 2\pi)/2 = -\pi$. This is the well-known half-quantum vortex of the polar phase~\cite{Autti2016}. When both $\theta$ and $\phi$ change by $\pi$, we end up back at the original state, guaranteeing the continuity of the total order parameter. From the form in Eq.~\eqref{eq:polarupdown} we can also immediately see that the superfluid state inside the vortex core where $\Delta_+ = 0$ corresponds to the $\beta$ phase of $^3$He with only $C_{-0}$:
\begin{equation}
    A_{\mu j} = \Delta_- e^{i\phi_-}(\hat{x}_\mu - i\hat{y}_\mu)\hat{z}_j.
    \label{eq:betaphase}
\end{equation}
The system remains in the superfluid state even in the core of the HQV.

We can also calculate the Ginzburg-Landau gradient energy using the form in Eq.~\eqref{eq:polarequivalency}, now with arbitrary $\hat{\bm{d}}$ and $\hat{\bm{m}}$:
\begin{align}
    f_{\text{grad}} &= K_1\nabla_k A^*_{\alpha j}\nabla_k A_{\alpha j} + K_2\nabla_j A^*_{\alpha j}\nabla_k A_{\alpha k} + K_3 \nabla_j A^*_{\alpha k}\nabla_k A_{\alpha j} \nonumber \\
    &= K_1\Delta^2\left(\vert\bm{\nabla}\phi\vert^2 + \vert\bm{\nabla}\theta\vert^2\right) + (K_2 + K_3)\Delta^2\left((\hat{\bm{m}}\cdot\bm{\nabla}\phi)^2 + (\hat{\bm{m}}\cdot\bm{\nabla}\theta)^2\right) \label{eq:polargrad}\\
    &= \frac{1}{2}K_1\Delta^2\left(\vert\bm{\nabla}\phi_+\vert^2 + \vert\bm{\nabla}\phi_-\vert^2\right) + \frac{1}{2}(K_2 + K_3)\Delta^2\left((\hat{\bm{m}}\cdot\bm{\nabla}\phi_+)^2 + (\hat{\bm{m}}\cdot\bm{\nabla}\phi_-)^2\right) \nonumber
\end{align}
where $K_1$, $K_2$, and $K_3$ are equal in the weak-coupling regime. Notably there are no terms in the gradient energy involving both $\phi_+$ and $\phi_-$. This means that in the traditional Ginzburg-Landau formalism with second order gradient terms, the vortices in different spin components don't interact at all until the cores start overlapping. In order to correctly account for the repulsion of vortices in different spin components, Fermi-liquid corrections are needed. This modification is discussed by Nagamura and Ikeda~\cite{Nagamura2018}, and it involves including the fourth order gradient terms in the Ginzburg-Landau energy expression.

\section{Double-core vortex in the B phase}\label{sec:doublecore}

As discussed earlier, the B phase is a state with zero total angular momentum, with only $C_{+-}$, $C_{00}$ and $C_{-+}$ components. This suggests a form similar to the polar phase Eq.~\eqref{eq:polarupdown} is possible for the B phase. Choosing the quantization axis along the spin-orbit rotation axis $\hat{\bm{\theta}}$, we can write Eq.~\eqref{eq:bphase1} equivalently as
\begin{equation}
    A_{\mu j} = \Delta_+ e^{i\phi_+}(\hat{d}_\mu + i\hat{e}_\mu)(\hat{m}_j - i\hat{n}_j) + \Delta_- e^{i\phi_-}(\hat{d}_\mu - i\hat{e}_\mu)(\hat{m}_j + i\hat{n}_j) + \Delta_0 e^{i\phi_0} \hat{f}_\mu \hat{l}_j
    \label{eq:bphaseupdown0}
\end{equation}
with the subscripts $+$, $-$, and $0$ referring to the spin projections of the Cooper pairs, and $\hat{\bm{f}} = \hat{\bm{d}}\times\hat{\bm{e}}$ and $\hat{\bm{l}} = \hat{\bm{m}}\times\hat{\bm{n}}$ pointing along the quantization axis $\hat{\bm{\theta}}$.

Similarly to Eq.~\eqref{eq:polarequivalency}, the two forms in Eqs.~\eqref{eq:bphase1} and \eqref{eq:bphaseupdown0} can be shown to be equivalent in the bulk with $\Delta_+ = \Delta_- = \Delta_0/2 = \Delta/2$:
\begin{align}
    A_{\mu j} &= \Delta_+ e^{i\phi_+}(\hat{d}_\mu + i\hat{e}_\mu)(\hat{m}_j - i\hat{n}_j) + \Delta_- e^{i\phi_-}(\hat{d}_\mu - i\hat{e}_\mu)(\hat{m}_j + i\hat{n}_j) + \Delta_0 e^{i\phi_0} \hat{f}_\mu \hat{l}_j \nonumber \\
    &= \Delta\hat{m}_j e^{i(\phi_+ + \phi_-)/2}\left[\cos\left(\frac{\phi_- - \phi+}{2}\right)\hat{d}_\mu + \sin\left(\frac{\phi_- - \phi_+}{2}\right)\hat{e}_\mu\right] \nonumber \\
    &+ \Delta\hat{n}_j e^{i(\phi_+ + \phi_-)/2}\left[\cos\left(\frac{\phi_- - \phi+}{2}\right)\hat{e}_\mu - \sin\left(\frac{\phi_- - \phi_+}{2}\right)\hat{d}_\mu\right] \nonumber \\
    &+ \Delta e^{i\phi_0}\hat{f}_\mu \hat{l}_j \nonumber \\
    &= \Delta \hat{m}_j e^{i\phi}\left(\cos\theta\hat{d}_\mu + \sin\theta\hat{e}_\mu\right) + \Delta \hat{n}_j e^{i\phi}\left(\cos\theta\hat{e}_\mu - \sin\theta\hat{d}_\mu\right) + \Delta e^{i\phi_0} \hat{f}_\mu \hat{l}_j \nonumber \\
    &= \Delta e^{i\phi} R(\bm{\theta})
    \label{eq:bequivalency}
\end{align}
with $\phi = \frac{\phi_+ + \phi_-}{2}$ and $\theta = \frac{\phi_- - \phi_+}{2}$, and the last step requiring that $\phi_0 = \phi$. States with $\phi_0 \neq \phi$ are also valid and can appear in the vortex cores, but they no longer represent the B phase. The assumption of fixed $\hat{\bm{\theta}}$ is quite limiting as it is known that most vortex structures involve more complicated spin textures in the "soft core" surrounding the vortex~\cite{Laine2018,Rantanen2024}, but we will see that it works well for the particular case of the double-core vortex of the B phase.

From Eq.~\eqref{eq:bequivalency} we can again consider possible vortex structures in terms of the winding of the different components far from the core. Far from the core the system is in the bulk B phase, $\phi_0 = \phi$. Parallel $2\pi$ winding in $\phi_+$ and $\phi_-$ results in the single-quantum phase vortex, with $\phi_0$ also winding by $2\pi$. Opposite winding in $\phi_+$ and $\phi_-$ results in a $\theta$ texture around the core, the spin vortex. The spin vortex has no winding in $\phi_0$, and in the core where $\Delta_+ = \Delta_- = 0$, the remaining $\Delta_0$ component corresponds to the polar phase, Eq.~\eqref{eq:polarphase}. In this representation, these structures are qualitatively similar to the phase and spin vortices in the polar phase, at least far from the vortex axis.

Unlike the polar phase, the B phase also supports a combination of the two structures, the spin-mass vortex~\cite{Kondo1992}. In the spin-mass vortex, $\phi_+$ (or $\phi_-$) winds by $4\pi$ while the other component has no winding. In this structure both $\phi$ and $\theta$ wind by $2\pi$ around the vortex core, a combination of a phase and spin vortex. In the polar phase this structure would be unstable and split into two HQVs, but as we will see the presence of the $\Delta_0$ component makes such a split unfavored in the B phase.

The presence of the $\Delta_0$ component in the B phase makes HQVs, $2\pi$ winding in only $\phi_+$ or $\phi_-$, distinct from their polar phase counterparts. While the $\pi$ winding of $\phi$ is compensated by $\theta$ for the up/down components, the $\pi$ winding of $\phi_0$ is not. Thus the $\Delta_0$ amplitude must go to zero on a line extending out from the vortex core. Suppression of $\Delta_0$ results in a planar phase wall, which is energetically very expensive and makes solitary HQVs unviable in the B phase.

Two HQVs, one each in $\phi_+$ and $\phi_-$, can form a pair connected by the planar wall, analogous to a Kibble-Lazarides-Shafi wall~\cite{Makinen2023}. This HQV pair is usually considered as a model of the double-core vortex of the B phase. As the planar wall costs a lot of energy, the separation between the pair is minimized. We show that the bulk double-core vortex is more accurately described by a bound state of three vortices, one in each spin component, instead of as a wall connecting two HQVs.

Detailed numerical calculations of the full structure of the double-core vortex~\cite{Regan2020,Rantanen2024} show that the suppressed component, $\Delta_0$, in the planar wall is the one in the direction pointing from one HQV to the other in Cartesian coordinates. If we take the vortex axis to be along $\hat{\bm{z}}$ and the two HQVs located on the $y$ axis, this suggests that the appropriate choice of quantization axis is $\hat{\bm{f}} = \hat{\bm{l}} = \hat{\bm{y}}$. In this basis the order parameter for a double-core vortex can be written as
\begin{equation}
    A_{\mu j} = \Delta_+ e^{i\phi_+}(\hat{z}_\mu + i\hat{x}_\mu)(\hat{z}_j - i\hat{x}_j) + \Delta_- e^{i\phi_-}(\hat{z}_\mu - i\hat{x}_\mu)(\hat{z}_j + i\hat{x}_j) + \Delta_0 e^{i\phi_0} \hat{y}_\mu \hat{y}_j
    \label{eq:doublecorevortex}
\end{equation}
with
\begin{align}
    \Delta_\pm &= \frac{\Delta}{2}|\tanh(\sqrt{x^2 + (y\mp a)^2}/2\xi)| \nonumber \\
    \Delta_0 &= \Delta |\tanh(\sqrt{x^2 + by^2}/2\xi)| \nonumber \\
    \phi_\pm &= \tan^{-1}((y\mp a)/x) \nonumber \\
    \phi_0 &= \tan^{-1}(y/x)
    \label{eq:doublecorevortex2}
\end{align}
where $2a$ is the spacing between the HQV cores, $\xi$ is the temperature and pressure dependent Ginzburg-Landau coherence length, and $b < 1$ sets the asymmetry of the vortex core. Numerical results suggest that $b \approx 0.5$ for the bulk double-core vortex.

We can calculate the kinetic energy of the double-core vortex using the Ginzburg-Landau gradient energy as in Eq.~\eqref{eq:polargrad}, without any fixed orientation for the basis:
\begin{align}
    f_{\text{grad}} &= K_1\nabla_k A^*_{\alpha j}\nabla_k A_{\alpha j} + K_2\nabla_j A^*_{\alpha j}\nabla_k A_{\alpha k} + K_3 \nabla_j A^*_{\alpha k}\nabla_k A_{\alpha j} \nonumber \\
    &= K_1\left(4\Delta_+^2|\bm{\nabla}\phi_+|^2 + 4\Delta_-^2|\bm{\nabla}\phi_-|^2 + \Delta_0^2|\bm{\nabla}\phi_0|^2\right) \nonumber \\
    &+ 2(K_2 + K_3)\Delta_+^2\left((\hat{\bm{n}}\cdot\bm{\nabla}\phi_+)^2 + (\hat{\bm{m}}\cdot\bm{\nabla}\phi_+)^2\right) \nonumber \\
    &+ 2(K_2 + K_3)\Delta_-^2\left((\hat{\bm{n}}\cdot\bm{\nabla}\phi_-)^2 + (\hat{\bm{m}}\cdot\bm{\nabla}\phi_-)^2\right) \nonumber \\
    &+ (K_2 + K_3)\Delta_0^2(\hat{\bm{l}}\cdot\bm{\nabla}\phi_0)^2.
    \label{eq:bphasegrad}
\end{align}
Again there are no terms involving a combination of $\phi_\pm$ or $\phi_0$, which means that traditional Ginzburg-Landau theory does not correctly account for the repulsion of vortices in separate spin components. The distance between the two HQV cores is mainly determined by the energy of the planar phase wall connecting them. When the length of the wall is on the order of the coherence length, further decrease in size is prevented by a combination of the logarithmically divergent kinetic energy and the complicated energetics of the vortex core itself.

\begin{figure}
    \centering
    \includegraphics[width=\linewidth]{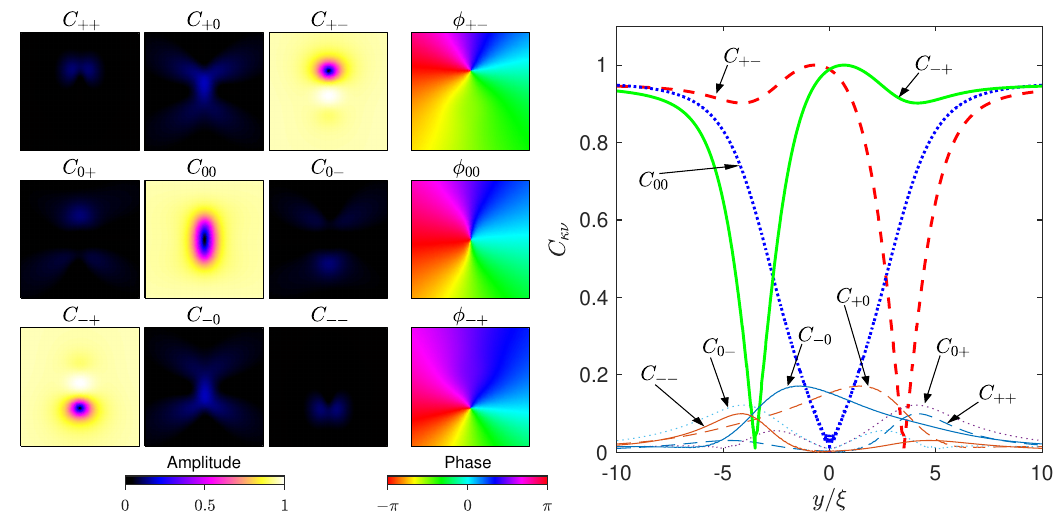}
    \caption{A double-core vortex with the two half-cores aligned along the $y$ axis. The data is taken from a numerical calculation at $p = 20$\,bar and $T = 0.80T_{\rm{c}}$. Distances are given in units of the temperature and pressure dependent Ginzburg-Landau coherence length $\xi\approx 30\text{ nm}$. (Left) The spatial profile of the order parameter in the angular momentum basis in the region $x,y\in [-10\xi, 10\xi]$. The amplitudes are shown for all nine components and the phase for the three components with phase winding in the bulk. The three distinct subcores are clearly visible in the $C_{-+}$, $C_{00}$, and $C_{+-}$ components. (Right) The same order parameter amplitudes on the $y$ axis. The half-cores are located at $y = \pm 3.5\xi$.}
    \label{fig:doublecorecomponents}
\end{figure}

We numerically calculate the structure of the double-core vortex in bulk $^3$He-B. The calculation method is described in Refs.~\cite{Rantanen2024,Rantanen2025}. Figure~\ref{fig:doublecorecomponents} shows the calculated order parameter of the double-core vortex represented in the angular momentum basis with the quantization axis taken as $\hat{\bm{y}}$. This representation reveals the distinct cores of the half-quantum vortices in the spin up ($C_{+-}$) and spin down ($C_{-+}$) components, with an apparent third core in the spin 0 ($C_{00}$) component located between them, matching the model given in Eqs.~\eqref{eq:doublecorevortex} and \eqref{eq:doublecorevortex2}. The triple-core structure is also clearly visible in the phase windings of the corresponding components, which are centered around three separate points. Interestingly in the bulk double-core vortex the $C_{00}$ component goes to zero only at a point at the center of the vortex, instead of in a line extending between the two cores, which distinguishes it from the model of a Kibble-Lazarides-Shafi wall connecting two HQVs.

The numerical result shows that components other than $C_{+-}$, $C_{00}$, and $C_{-+}$ become nonzero near the cores. These components appear in order for the state to minimize the complicated bulk energy of $^3$He, and are required to explain some features of the double-core vortex, such as the core magnetization. In the simple model of Eq.~\eqref{eq:doublecorevortex} the two HQV cores have magnetization in the $\pm\hat{\bm{y}}$ direction that cancel each other, but no magnetization along $\hat{\bm{z}}$, unlike the numerical result~\cite{Thuneberg1987,Regan2020,Rantanen2024}.

The order parameter state between the two HQV cores located at $y = \pm 3.5\xi$ deviates from the bulk B phase, as $\phi_0 \neq \phi$. Precisely between the two cores at $y  = 0$ the third component goes to zero, $\Delta_0 = 0$, and the order parameter matches that of the planar phase. When $y > 0$ ($y < 0$), the value of $\phi_0$ is $\pi/2$ ($-\pi/2$) different from $\phi = \frac{\phi_+ + \phi_-}{2} = 0$. This state does not correspond to any of the known inert phases~\cite{Barton1974}. In a simple model of the double-core vortex, involving only the major $C_{-+}$, $C_{00}$ and $C_{+-}$ components, the state between the HQVs can be written as
\begin{equation}
    A_{\mu j} = -i\Delta_+ (\hat{z}_\mu + i\hat{x}_\mu)(\hat{z}_j - i\hat{x}_j) + i\Delta_-(\hat{z}_\mu - i\hat{x}_\mu)(\hat{z}_j + i\hat{x}_j) \pm i\Delta_0 \hat{y}_\mu \hat{y}_j
\end{equation}
with the upper sign corresponding to $y > 0$ and the lower to $y < 0$. At the cores of the two HQVs where one of the $\Delta_\pm = 0$, the order parameter is non-unitary and looks like a linear combination of the order parameters of the polar and A$_1$ phases~\cite{VollhardtWolfle}. The non-unitarity of the cores is the reason for the magnetization of the double-core vortex. Outside the HQV core, when $|y| > 3.5\xi$, the phases once again match ($\phi_0 = \phi$) and the order parameter can smoothly deform to the bulk B phase. This deformed intermediate state resembles B phase in a magnetic field, referred to as the B$_2$ phase~\cite{VollhardtWolfle}.

The double-core vortex shown in Figure~\ref{fig:doublecorecomponents} is observed in bulk $^3$He-B and corresponds to tightly bound pair of HQVs. Experiments in anisotropic confinement show that HQVs with larger separation can be stabilized in the polar-distorted B phase (pdB) by cooling from polar phase with HQVs~\cite{Makinen2019}. The HQVs are believed to be pinned on the nanoscale features of the confining material, preventing them from approaching each other. We numerically calculate the state of the stretched double-core vortex in bulk B phase, simulating the pinning sites by cylindrical objects with a diameter of the coherence length $\xi$. We place a single pinning site inside each HQV core of the double-core vortex, and calculate the state for different distance between the two sites. The order parameter form in Eq.~\eqref{eq:bphaseupdown0} is used as a convenient initial condition, and the minimum energy structure for the full $3\times 3$ order parameter is found numerically. The pinning sites are modeled as physical objects, with specular boundary conditions at the boundary, so that the perpendicular orbital components (in Cartesian basis) go to zero at the wall and the parallel components have zero gradients through the boundary.

\begin{figure}
    \centering
    \includegraphics[width=\linewidth]{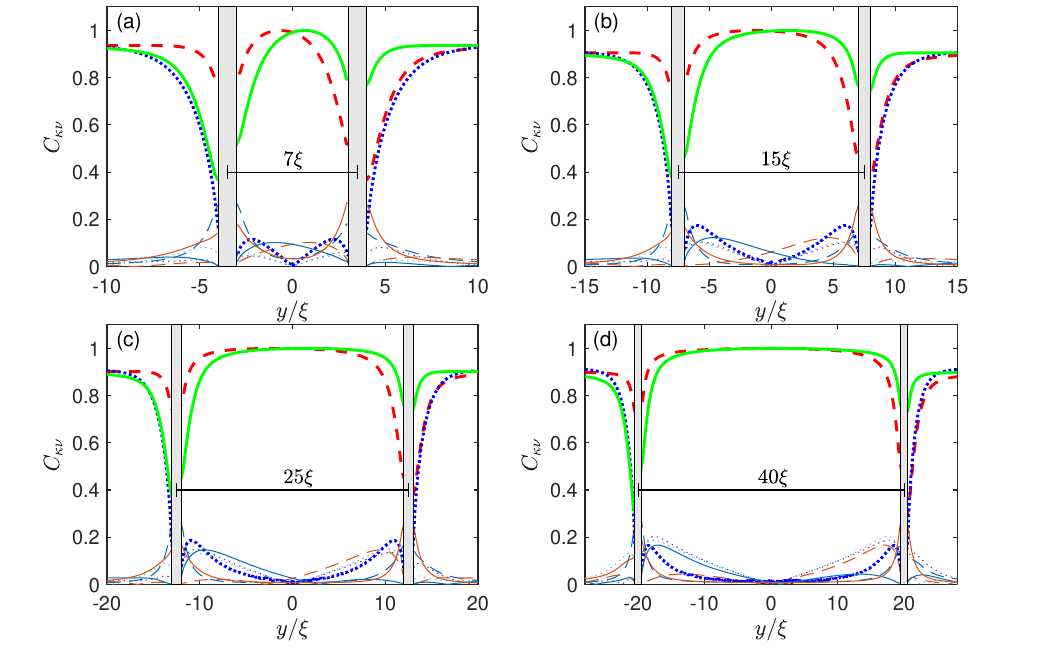}
    \caption{The order parameter along the $y$ axis for a stretched double-core vortex calculated at $p = 20\text{ bar}$ and $T = 0.80T_{\rm{c}}$. The lines represent the components as labeled in Figure~\ref{fig:doublecorecomponents}. The pinning sites are cylindrical obstacles with a diameter of $\xi$, marked by the grey regions. The state shown in panel (a) corresponds to $7\xi$ distance between cores, which is the equilibrium distance found in the bulk. Panels (b), (c) and (d) correspond to spacings of $15\xi$, $25\xi$ and $40\xi$, respectively.}
    \label{fig:stretchedcomponents}
\end{figure}

Figure~\ref{fig:stretchedcomponents} shows the results of our calculations. The effect of pinning sites without stretching was tested by placing the sites $7\xi$ apart, corresponding to the distance between HQVs in the bulk. The order parameter, shown in Figure~\ref{fig:stretchedcomponents}a, is similar to the bulk case in Figure~\ref{fig:doublecorecomponents}, with the main difference being the behavior of the $C_{00}$ component near the HQV cores. In the case with pinning sites, the $C_{00}$ component goes to zero both at $y = 0$, but also near the cores of the two HQVs.

As the two cores are stretched further apart, the state between them changes qualitatively from a single point of planar phase at $y = 0$ to a full line where $C_{00}$ remains zero. Near the pinning sites, the structure remains similar to the unstretched case. Only at higher separations does the description of two HQVs connected by a planar wall become accurate, while in the tightly bound bulk case the structure more accurately corresponds to a composite of three distinct cores. It is notable that even the relatively small structure is enough to strongly pin the vortex, and the increased energy from the longer planar wall is not enough to depin the HQVs. The depinning force from the planar wall is not expected to increase as the cores get further apart, as the energy density of the wall is independent of the separation for large distances. This supports the experimental observation of strong pinning of vortices in anisotropic confinement by nanostrands~\cite{Makinen2019}.

\section{Discussion and Conclusion}\label{sec:conclusion}

We present a simple analytic form for the order parameter of the double-core vortex in superfluid $^3$He-B. The division of the B phase order parameter into three separate spin components allows for easy classification of possible vortex structures, including the spin-mass and half-quantum vortices. We represent the double-core vortex as three separate vortices in the three spin components of the B phase, in contrast to the traditional description of combination of two half-quantum vortices.

Numerical calculation of the double-core vortex structure supports the division into three distinct cores. The length of the planar wall connecting the two HQVs is minimized in the bulk, so that the planar order parameter is achieved only in a single point in the middle of the structure. In this state the double-core vortex truly appears to be a bound state of three separate vortices. In experiments involving anisotropic confinement, the HQVs are expected to be separated by a considerable amount due to pinning of the cores on confining structures. We simulate the stretched double-core vortex and show that there is a qualitative change from a single point of planar phase to a line when the separation between cores becomes large enough.

We note that the traditional Ginzburg-Landau formalism does not correctly account for the repulsion between half-quantum vortices, and Fermi-liquid corrections are needed. Indeed, calculations using quasiclassical theory suggest a much larger separation between subcores for the double-core vortex~\cite{Silaev2015}. However, quasiclassical theory does not account for strong-coupling corrections of the bulk energy, which can be important for the structure of the core. A full simulation of the double-core vortex including all the relevant effects remains a task for the future. While these details may prove to be important for quantitative comparison with experiments, the major qualitative features of the double-core vortex can be intuitively understood through our simple model. 

\backmatter

\bmhead{Acknowledgements}
We thank Vladimir Eltsov and Erkki Thuneberg for useful discussions. We acknowledge the computational resources provided by the Aalto Science-IT project.

\bibliography{sn-bibliography}

\end{document}